\documentclass[bibnotes,showpacs,twocolumn,aps,prl,a4paper]{revtex4}

\usepackage{color,epsfig,rotating,graphicx,subfigure}
\usepackage{amsmath,amssymb,amscd}

\usepackage[latin1]{inputenc}
\usepackage[english]{babel}

\usepackage{dsfont}
\usepackage{textcomp}

\begin{document}

%
%
\title{On the limits of the effective description of hyperbolic materials in presence of surface waves}


\author{Maria Tschikin and Svend-Age Biehs$^{*}$}

\address{
Institut f\"{u}r Physik, Carl von Ossietzky Universit\"{a}t, D-26111 Oldenburg, Germany\\
$^*$Corresponding author: s.age.biehs@uni-oldenburg.de
}

\author{Riccardo Messina and Philippe Ben-Abdallah}

\address{
Laboratoire Charles Fabry, UMR 8501, Institut d'Optique, CNRS, Universit\'{e} Paris-Sud 11, 2, Avenue Augustin Fresnel, 91127 Palaiseau Cedex, France 
}

\date{\today}

\pacs{73.20.Mf, 78.67.Pt}

\begin{abstract}
Here, we address the question of the validity of an effective description for hyperbolic metamaterials in the near-field region. We show that the presence of localized modes such as surface waves drastically limits the validity of the effective description and requires revisiting the concept of homogenization in near-field. We demonstrate from exact calculations that one can find surface modes in spectral regions where the effective approach predicts hyperbolic modes only. Hence, the presence of surface modes which are not accounted for in the effective description can lead to physical misinterpretations in the description of hyperbolic materials and their related properties.
\end{abstract}

\maketitle
\newpage

%
%
With today's nanotechnology it is possible to manufacture 
artificial composite materials at tiny scales which 
manifest optical properties that we do not encounter in nature. 
These media also called metamaterials are usually structured 
at the length scale or below the wave length of photons. 
These media have been largely exploited during the last 
decades to overthrow long standing paradigms in physics. For 
instance, new phenomena such as the negative refraction~\cite{Smith2000}, the 
super-resolution~\cite{Pendry2000} or reversed Doppler effects~\cite{Joannopoulos2003} were 
predicted for metamaterials. 

One class of metamaterials, the so called indefinite or hyperbolic 
materials (HM) has recently attracted much attention. 
In fact, the dispersion relation of electromagnetic waves in these media can be represented by hyperbolic isofrequency curves meaning that they support 
propagating modes having extremely large wavevectors far beyond the light line~\cite{SmithSchurig2003,HuChui2002}. 
Because of that property HMs are viewed as promising candidates to 
develop or improve breakthrough technologies among which are the 
sub-diffractive imagery (hyperlensing)~\cite{JacobEtAl2006,SalandrinoEngheta2006,LiuEtAl2007}, 
the thermal management at nanoscale~\cite{BiehsEtAl2012,GuoEtAl2012,NarimanovSmolyaninov2011}, 
the near-field energy conversion~\cite{Nefedov2011}, enhanced light 
emission~\cite{ZubinEtAl2012}, and quantum information systems~\cite{PoddubnyEtAl2011}.

The exotic properties of HMs in the near-field regime are 
conventionally investigated within the framework of the effective medium 
theory (EMT). Naturally the question arises under which conditions the 
EMT remains valid in the close vicinity of a composite material.  In general 
it is believed that for propagating modes the EMT is valid as long as the 
wavelength is much larger than the size of the unit-cell of the composite material. 
In the evanescent regime this condition imposes that
the wavevector of the evanescent field has to be smaller than
the inverse of the unit-cell size as well~\cite{ChebykinEtAl2012}. 
In the context of the spontaneous emission it was shown that the EMT can greatly 
underestimate or overestimate~\cite{KidwaiEtAl2012,IorshEtAl2012,KidwaiEtAl2011} 
the Purcell factor when this condition is not met. In addition,
when the composite materials support surface waves one encounters 
a nonlocality which cannot be adequately described by the 
EMT~\cite{ChebykinEtAl2012,OrlovEtAl2011}. Another nonlocal effect
can arise from the nonlocal response of the composite materials
for very large wave vectors~\cite{YanEtAl2012}. Therefore, a nonlocal
effective description is needed to take such effects into account~\cite{ElserEtAl2007}. 

In this letter, we show that the presence of surface waves at the interface 
of the basic components of HMs also requires revisiting the concept of homogenization 
which otherwise, in the EMT framework, can lead to quantitatively wrong and 
physically misinterpreted results if one estimates the hyperbolic bands from
EMT~\cite{Krishnamoorthy2012,Noginov2012}. By comparing the exact calculation 
of the local density of states of electromagnetic field at arbitrary distances 
above an HM with the predictions given by the EMT we highlight the regions 
where the effective theory fails to describe properly the field outside the HM. 
In particular, we demonstrate from exact calculations the existence of surface modes inside the hyperbolic bands which are not
predicted at all by the EMT. Hence, some properties attributed to the hyperbolicity of material could be misinterpreted or even worse could not exist anymore. For example large changes in the Purcell factors within the hyperbolic frequency bands could be due to such surface modes, which are not taken into account in the EMT. Hence, the EMT would in this case give results which are quantitatively and qualitatively wrong. 

\begin{figure}[htb]
\centerline{\includegraphics{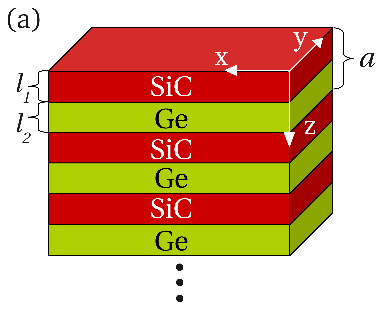} \includegraphics{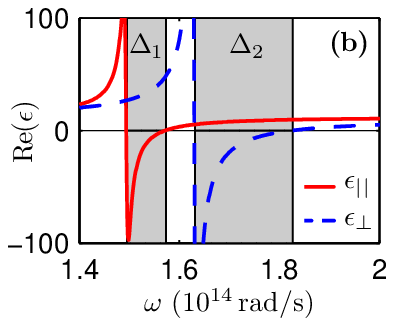}}
\caption{(Color online) (a) Sketch of the 1D SiC/Ge periodic structure with their layer thicknesses $l_1$ and $l_2$. The period of the structure is $a=l_1+l_2$ and the number of layers is equal to $N=100$. The surrounding medium is vacuum. (b) Real part of the effective permittivities $\epsilon_{||}$ and $\epsilon_{\perp}$ for $f=0.4$. The grey areas $\Delta_1$ and $\Delta_2$ mark the hyperbolic regions where Re($\epsilon_{||}$)Re($\epsilon_{\perp})<0$.}
\label{Fig:LayeredMedium}
\end{figure}

To estimate the range of validity for the EMT in the presence of surface modes we consider a 1D periodic structure composed by two materials with one which naturally supports surface waves in the spectral range of interest. This structure is depicted in Fig.~\ref{Fig:LayeredMedium}(a). One basis material is Silicon carbide (SiC) and the other one is germanium (Ge). The polar medium SiC supports surface phonon polariton (SPP) modes in the mid-infrared at $\lambda_{\rm SPP}=10.3\,\mu$m and its permittivity is given by 
\begin{equation}
\epsilon_{\rm SiC}     = \epsilon_{\infty}\bigl(\frac{\omega_{\rm L}^2 -\omega^2 -\rm i \gamma \omega}{\omega_{\rm T}^2 -\omega^2 -\rm i \gamma \omega}\bigr)
\end{equation}
following the Drude Lorenz model~\cite{Palik} with $\epsilon_{\infty}=6.7$, $\omega_{\rm L}=182.7\cdot 10^{12}\,$rad/s, $\omega_{\rm T}=149.5\cdot 10^{12}\,$rad/s and $\gamma=0.9\cdot 10^{12}\,$rad/s. The dielectric permittivity of Ge in the same frequency range is constant and equal to $\epsilon_{\rm Ge}=16$. According to the EMT, for $a<\lambda_{\rm SPP}$ the equivalent homogenized medium to this structure is an uniaxial anisotropic medium with an effective permittivity parallel and perpendicular  to the optical axis given by the expressions~\cite{Cortes2012} 
\begin{equation}
 \epsilon_{||}=f\epsilon_{\rm SiC} + (1-f) \epsilon_{\rm Ge} \ \ \ \ \ \text{and} \ \ \ \ \ \epsilon_{\perp} = \frac{\epsilon_{\rm SiC}\epsilon_{\rm Ge}}{f\epsilon_{\rm Ge} + (1-f) \epsilon_{\rm SiC}}
\label{EQ:Permittivitaet} 
\end{equation}                                                      
with the volume filling fraction $f=l_1/a$. The reflection coefficients for s- and p-polarized waves in the effective description are thus 
\begin{eqnarray}
 r_{\rm s} = \frac{k_{z0}-k_{\rm s}}{k_{z0}+k_{\rm s}}  \ \ \ \ \ \text{and} \ \ \ \ \  r_{\rm p} = \frac{\epsilon_{||}k_{z0} - k_{\rm p}}{\epsilon_{||}k_{z0} + k_{\rm p}}
\label{EQ:RefKoefEFF}
\end{eqnarray}
with $k_{z0}=\sqrt{\omega^2/c^2-\kappa^2}$, $k_{\rm s}=\sqrt{\omega^2/c^2\epsilon_{||}-\kappa^2}$ and $k_{\rm p}=\sqrt{\omega^2/c^2\epsilon_{||}-\kappa^2\frac{\epsilon_{||}}{\epsilon_{\perp}}}$ the normal components of the wavevector in vacuum and in the medium for s and p waves. Here the wave vector $\vec\kappa = (k_x,k_y)^{\rm t}$ is perpendicular to the optical axis. Therefore p-polarized electromagnetic waves (the so called extra-ordinary waves) in such an effective uniaxial medium  fulfill the dispersion relation~\cite{Yeh}
\begin{equation}
 \frac{\kappa^2}{\epsilon_{\perp}} + \frac{k_z^2}{\epsilon_{||}} = \frac{\omega^2}{c^2}.
\label{EQ:HyperbDisp}
\end{equation}
If Re($\epsilon_{||}$) and Re($\epsilon_{\perp}$) are both positive Eq.~(\ref{EQ:HyperbDisp}) describes an elliptic dispersion curve in the ($\kappa$,$k_z$) plane while on the contrary if the effective permittivities have opposite signs, i.e., if Re($\epsilon_{||}$)Re($\epsilon_{\perp})<0$, the dispersion curve described by Eq.~(\ref{EQ:HyperbDisp}) becomes hyperbolic. In the spectral range where Re($\epsilon_{||}$)Re($\epsilon_{\perp})<0$ waves are propagating and are called hyperbolic modes. In Fig.~\ref{Fig:LayeredMedium}(b) Re($\epsilon_{||}$) and Re($\epsilon_{\perp}$) are plotted versus the frequency $\omega$ and the hyperbolic regions are highlighted in grey. For the chosen filling fraction $f=0.4$ the first band $\Delta_1$ represents the case where Re$(\epsilon_{||})<0$ and Re$(\epsilon_{\perp})>0$ whereas in the second band $\Delta_2$ the effective permittivities fulfill Re$(\epsilon_{||})>0$ and Re$(\epsilon_{\perp})<0$.    

To check the pertinence of the EMT predictions we use the exact S-matrix method~\cite{FrancoeurAPL,PBA2009,Yeh} with a finite but large number of periods $N/2$. Beside, for our calculations we choose $N=100$ layers where the last Ge layer is assumed to be infinitely large extending to $z \rightarrow \infty$. For $N \rightarrow \infty$ it is well known that propagating waves in a periodic structure are Bloch waves and satisfy the Bloch mode dispersion relation~\cite{Yeh}
\begin{equation}
\begin{split}
 \cos(k_za)&=-\frac{1}{2}\Bigl( \frac{P_{i2}k_{z1}}{P_{i1}k_{z2}} + \frac{P_{i1}k_{z2}}{P_{i2}k_{z1}} \Bigr) \sin(k_{z1}l_1) \sin(k_{z2}l_2)  \\
&\quad +\cos(k_{z1}l_1) \cos(k_{z2}l_2),
\end{split}
\label{EQ:BlochDisp}
\end{equation}
with $P_{{\rm s}1}=P_{{\rm s}2}=1$, $P_{{\rm p}1}=\epsilon_{\rm SiC}$ and $P_{{\rm p}2}=\epsilon_{\rm Ge}$.

\begin{figure}[htb]
\centerline{\includegraphics{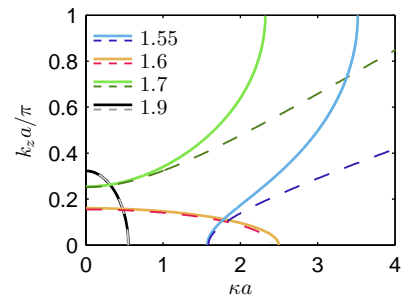}}
\caption{(Color online) Isofreqency curves for the Bloch dispersion relation~(\ref{EQ:BlochDisp}) (solid lines) and the effective dispersion relation~(\ref{EQ:HyperbDisp}) (dashed lines) for $f=0.4$, $a=500\,$nm and for $\omega=1.55\cdot10^{14}\,$rad/s, $1.6\cdot10^{14}\,$rad/s, $1.7\cdot10^{14}\,$rad/s, and $1.9\cdot10^{14}\,$rad/s.}
\label{Fig:Contourplot}
\end{figure}

To analyse the dispersion relation for the exact and the effective case we have plotted  the isofrequency curves in the $\kappa$-$k_z$ plane from Eq.~(\ref{EQ:HyperbDisp}) and (\ref{EQ:BlochDisp}) for different frequencies in Fig.~\ref{Fig:Contourplot}. Here we used the Bloch Eq.~(\ref{EQ:BlochDisp}) which gives the exact isofrequency curves for the limiting case $N\rightarrow \infty$. For the frequencies $\omega=1.6\cdot10^{14}\,$rad/s and $\omega=1.9\cdot10^{14}\,$rad/s outside the hyperbolic regions  the effective permittivities are both positive as shown in Fig.~\ref{Fig:LayeredMedium}(b) and accordingly we get elliptical isofrequency curves. The difference between the principal axes of the black and orange curves stems from the fact that Re$(\epsilon_{||})<{\rm Re}(\epsilon_{\perp})$ for the frequency range between $\Delta_1$ and $\Delta_2$ whereas for frequencies larger (smaller) than $\Delta_2$ ($\Delta_1$) Re$(\epsilon_{||})>{\rm Re}(\epsilon_{\perp})$. The Bloch dispersion curves (solid line) for these frequency regions fit excellently the effective data (dashed line). 

The green and blue curves in Fig.~\ref{Fig:Contourplot} are plotted for frequencies which are located in the hyperbolic bands. The dashed lines for $\omega=1.55\cdot10^{14}\,$rad/s (blue) and $\omega=1.7\cdot10^{14}\,$rad/s (green) illustrate the hyperbolic effective results where the blue curve is an example for a dispersion curve from the $\Delta_1$ band and the green one represents a dispersion curve from the $\Delta_2$ band. Here the Bloch dispersion coincides with the EMT result for small $\kappa$. For increasing values of $\kappa$ the solid curves show the existence of a $\omega$-dependent vertical asymptote as a consequence of Bloch Eq.~(\ref{EQ:BlochDisp}), while the effective dashed curves do not have a maximal $\kappa$ value. This is a major difference between the behavior of the structure predicted by the effective theory and by the rigorous theory.     

\begin{figure}[htb]
\centerline{\includegraphics{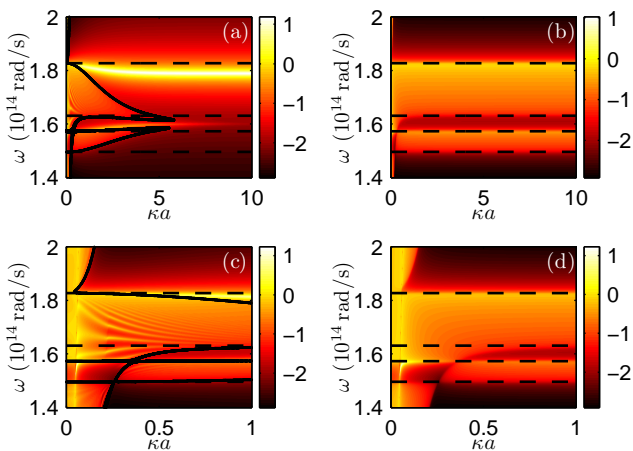}}
\caption{(Color online) Log of RC [$\kappa<\omega/c: 1 - |r_{\rm p}|^2$; $\kappa>\omega/c: {\rm Im}(r_{\rm p})$] for a SiC/Ge structure and two different $\kappa$ regimes with $a=100\,$nm, $f=0.4$ and $N=100$. (a), (c) show the exact and (b), (d) the effective case. The black solid lines indicate the boundaries of the Bloch mode dispersion relation from Eq.~(\ref{EQ:BlochDisp}). The black dashed lines mark the two hyperbolic bands $\Delta_1$ and $\Delta_2$.}
\label{Fig:Reflectioncoef_a100nm}
\end{figure}
      
Although the isofrequency curves deduced from the effective medium theory and from the Bloch theory coincide very well for $\kappa<1/a$ as it is well known, the exact calculations can deviate very strongly from the effective and Bloch results due to the fact that the structure can support surface modes which are not taken into account neither in the Bloch theory nor in the EMT. To demonstrate this fact we have  plotted in Fig.~\ref{Fig:Reflectioncoef_a100nm} both the exact and the effective reflection coefficients (RCs) in the $\omega$-$\kappa$ plane in  regions of small and large $\kappa$ values. In Fig.~\ref{Fig:Reflectioncoef_a100nm}(a) and (c), where the exact RC is plotted, the black solid lines represent the boundaries of the Bloch bands given by Eq.~(\ref{EQ:BlochDisp}) for propagating modes in the medium. Here the considered frequency range is still in the first Bloch band. Inside these Bloch areas the RC has large values and with increasing number $N$ these areas are filled due to an increasing number of discrete Bloch modes. The maximal $\kappa$ value for the contributing Bloch modes is therefore given by Eq.~(\ref{EQ:BlochDisp}).

The RC for the EMT is plotted in Fig.~\ref{Fig:Reflectioncoef_a100nm}(b) and (d). The black dashed lines mark the hyperbolic bands $\Delta_1$ and $\Delta_2$ where Re($\epsilon_{||}$)Re($\epsilon_{\perp})<0$. By comparing Fig.~\ref{Fig:Reflectioncoef_a100nm}(a) with Fig.~\ref{Fig:Reflectioncoef_a100nm}(b) with  $\kappa_{\rm max} = 10/a$ it can be seen that the EMT fails to reproduce the exact result for large $\kappa$ values, as already mentioned before. The boundaries for the Bloch bands change drastically  whereas the size of the hyperbolic bands does not change with increasing $\kappa$ values. In Fig.~\ref{Fig:Reflectioncoef_a100nm}(c) and (d) the exact and effective RC is plotted up to  $\kappa_{\rm max} = 1/a$. Here the RC caclulated with the EMT nearly coincides with the exact result. The exact calculation shows discrete bloch modes in the hyperbolic regime. This comes from the limited number of layers ($N = 100$). For $N\rightarrow \infty$ the discrete Bloch modes will fill the whole hyperbolic regime.

But beyond the Bloch areas in Fig.~\ref{Fig:Reflectioncoef_a100nm}(a) two prominent features around $\omega_1=1.8\cdot 10^{14}\,$rad/s (upper branch) and $\omega_2=1.6\cdot 10^{14}\,$rad/s (lower branch) are visible with very high values for the RC. These modes are coupled SPP modes stemming from the first SiC layer on the top of the structure. These surface modes are damped on a scale much smaller than the period $a$ so that they do not 'feel' the periodic structure. The upper branch around $\omega_1$ has a very high RC and moreover high values up to very large $\kappa$. The SPP contribution becomes therefore relevant for very small distances $z$ to the periodic structure. In the exact calculations, as it can be seen in Fig.~\ref{Fig:Reflectioncoef_a100nm}(a), the upper SPP branch is located in the second hyperbolic band $\Delta_2$ where the effective results do not show any SPP mode in the hyperbolic regimes. The contribution of this SPP mode around $\omega_1$ is still visible for $\kappa<1/a$ as it can be seen in Fig.~\ref{Fig:Reflectioncoef_a100nm}(c). The contribution of the coupled surface modes reduces but does not vanish. This qualitative difference between both approaches can lead to misinterpretations for very small distances to the structure. 

\begin{figure}[htb]
\centerline{\includegraphics{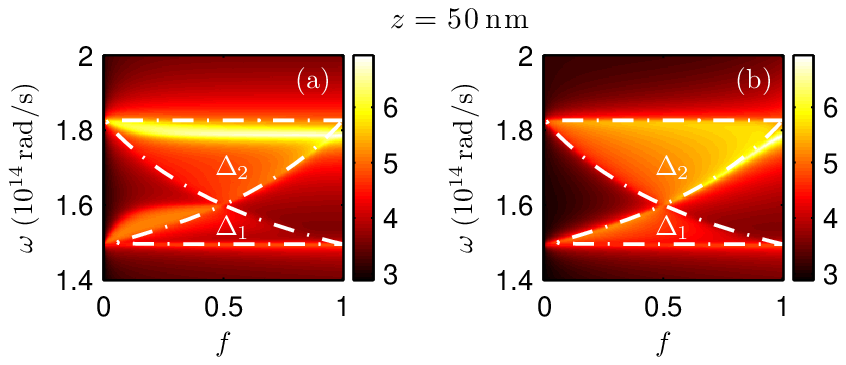}}
\centerline{\includegraphics{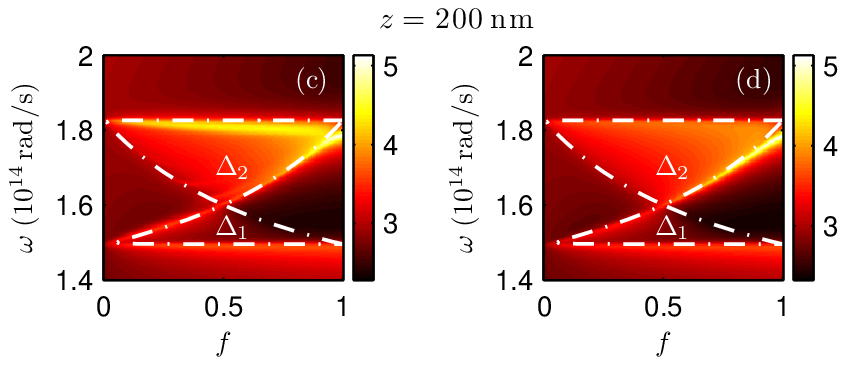}}
\centerline{\includegraphics{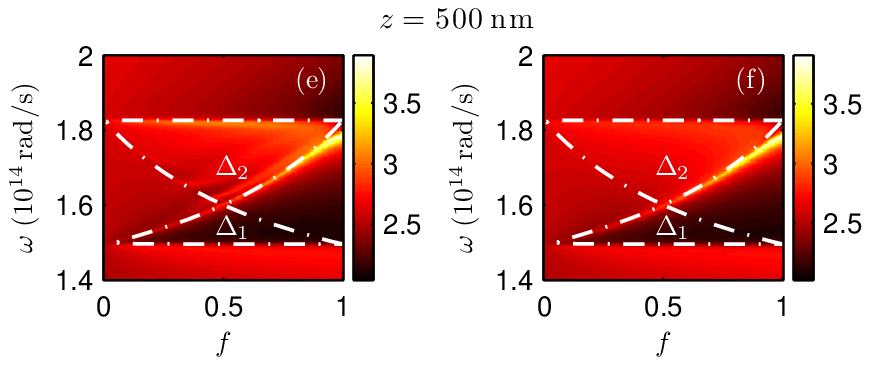}}
\caption{(Color online) Local Density of States (LDOS) in the $\omega$-$f$ plane for different distances $z$ above the SiC/Ge bilayer structure calculated with the exact (a), (c), (e) and the effective (b), (d), (f) theory with $a=100\,$nm and $N=100$. The dot-dashed white lines mark the hyperbolic regions $\Delta_1$ and $\Delta_2$.}
\label{Fig:LDOS}
\end{figure}

To estimate the range of validity of the EMT and to evaluate the disturbing role played by the SPP mode we have finally calculated the local density of states (LDOS) of electromagnetic field for all filling fractions at different distances $z$ above the SiC/Ge structure both for the homogenized structure and for the real one. The results are plotted in Fig. 4. For $z=500\,$nm in Fig.~\ref{Fig:LDOS}(e) and (f) for both plots high LDOS values appear for large $f$ at the lower right boundary of $\Delta_2$. This is due to a surface mode outside the hyperbolic bands where Re$(\epsilon_{||})$Re$(\epsilon_{\perp})=1$ and Re$(\epsilon_{||})<0$. These conditions together with Eq.~(\ref{EQ:Permittivitaet}) lead to $f>0.5$ as a necessary condition for having surface modes outside the hyperbolic band within the EMT. The contribution of the coupled SPP mode of the first layer within the hyperbolic band $\Delta_2$ is very low and the effective calculations fit very well the exact results. However, when decreasing the distance $z$ this contribution of the coupled SPP mode becomes larger and eventually for $z=50\,$nm the effective LDOS is not able to mimic the exact LDOS as illustrated in Fig.~\ref{Fig:LDOS}(a) and (b). Here the contribution of the coupled SPP mode of the first layer is much larger than the contribution of the Bloch waves. Hence, from the EMT calculations one would expect a large LDOS stemming from the hyperbolic modes, which would be a wrong conclusion, since as we see from the exact results the large LDOS is due to SPP modes from the first layer. 
 
In this work we have shown that even for distances larger than the lattice period the EMT does not provide a correct physical picture of the structure when localized modes such as surface polaritons are present. We have shown that the usual conditions for the validity of the EMT  $a \ll \lambda$ and $\kappa a \ll 1$ have to be augmented by the condition that the distance $z$ at which  the electromagntic field is investigated  must be larger than the penetration length of the SPP modes into the vacuum region.

%
%

\begin{acknowledgments}
M.T. gratefully acknowledges support from the Stiftung der
Metallindustrie im Nord-Westen. The authors acknowledge financial support by the DAAD and
Partenariat Hubert Curien Procope Program (project 55923991).
\end{acknowledgments}

%
%

\appendix

%
%

\end{document}